\documentclass[twocolumn,showpacs,amsmath,nofootinbib,pra,aps,amssymb,longbibliography]{revtex4-2}

\usepackage[T1]{fontenc}
\usepackage[utf8]{inputenc}
\usepackage{times}
\usepackage{color} 
\usepackage{array}
\usepackage{amssymb,amsmath}
\usepackage{amsbsy}
\usepackage[pdftex]{graphicx} 
\usepackage{bm}
\usepackage{float}
\usepackage{dcolumn}
\usepackage{scalerel}
\usepackage{tikz}
\usetikzlibrary{svg.path}

\definecolor{orcidlogocol}{HTML}{A6CE39}
\tikzset{
  orcidlogo/.pic={
    \fill[orcidlogocol] svg{M256,128c0,70.7-57.3,128-128,128C57.3,256,0,198.7,0,128C0,57.3,57.3,0,128,0C198.7,0,256,57.3,256,128z};
    \fill[white] svg{M86.3,186.2H70.9V79.1h15.4v48.4V186.2z}
                 svg{M108.9,79.1h41.6c39.6,0,57,28.3,57,53.6c0,27.5-21.5,53.6-56.8,53.6h-41.8V79.1z M124.3,172.4h24.5c34.9,0,42.9-26.5,42.9-39.7c0-21.5-13.7-39.7-43.7-39.7h-23.7V172.4z}
                 svg{M88.7,56.8c0,5.5-4.5,10.1-10.1,10.1c-5.6,0-10.1-4.6-10.1-10.1c0-5.6,4.5-10.1,10.1-10.1C84.2,46.7,88.7,51.3,88.7,56.8z};
  }
}

\newcommand\orcidicon[1]{\href{https://orcid.org/#1}{\mbox{\scalerel*{
\begin{tikzpicture}[yscale=-1,transform shape]
\pic{orcidlogo};
\end{tikzpicture}
}{|}}}}

\usepackage[unicode,breaklinks]{hyperref}
\hypersetup{
    unicode=true,
    plainpages=false, 
    colorlinks=true,
    linkcolor=blue,
    citecolor=blue,
    filecolor=black,
    urlcolor=blue
}
\usepackage{url}
\usepackage{verbatim}

\newcommand{\add}{a_\mathrm{dd}}

\newcommand{\br}{\mathbf{r}}
\newcommand{\bx}{\mathbf{x}}

\newcommand\gammaQF{\gamma_\mathrm{QF}}

\begin{document}
  
\title{Honeycomb supersolid --   Dirac points and shear-instability induced crystal transitions}
\author{P.~B.~Blakie\orcidicon{0000-0003-4772-6514}}
\affiliation{%
	Dodd-Walls Centre for Photonic and Quantum Technologies, Dunedin 9054, New Zealand}
\affiliation{Department of Physics, University of Otago, Dunedin 9016, New Zealand}
\date{\today}
\begin{abstract}   
The honeycomb supersolid state is predicted to form in a dipolar Bose-Einstein condensate with a planar confining potential. Our results for its excitation spectrum  reveal the  gapless  bands and the emergence of Dirac points at the Brillouin zone edge, {manifesting as points where the second sound and transverse sound bands touch}.  
The honeycomb supersolid has three sound speeds that we  connect to its  elastic parameters through hydrodynamic theory. From this analysis we find conditions where a shear instability occurs as the honeycomb rigidity disappears. This gives insight into the nonequilibrium dynamics following an interaction quench, where the honeycomb pattern melts and different crystal orders emerge.
\end{abstract}

\maketitle

A two-dimensional (2D) triangular supersolid has been realized using a dipolar Bose-Einstein condensate (BEC) confined in a trapping potential with an oblate shape \cite{Norcia2021a,Bland2022a}. This system exhibits a complex phase diagram, featuring various ground-state crystal patterns, separated by first-order phase transitions \cite{Zhang2019a,Poli2021a,Hertkorn2021b,Zhang2021a,Zhang2023a,Ripley2023a}. One of the intriguing phases that emerges is the honeycomb-patterned supersolid, which combines high modulational contrast with a significant superfluid fraction \cite{Zhang2019a,Gallemi2022a}. Experimental developments, such as the production of larger dipolar condensates \cite{Jin2023a,Krstajic2023a} and progress with polar molecule condensates \cite{Schmidt2022a,Bigagli2024a},
 should soon lead to the experimental realization of honeycomb supersolids.
 
Honeycomb lattices garnered considerable interest following studies on graphene \cite{Zhang2005a,Novoselov2005a}. In graphene  the band structure exhibits degeneracies where two bands touch at the corners of the first Brillouin zone, known as the Dirac points. Near the Dirac points, the excitation bands take on a cone-like shape, akin to massless relativistic particles. Cold-atom systems have been used to engineer similar band structures with the honeycomb structure imposed externally by optical lattices \cite{Grynberg1993a,Wu2007a,Zhu2007a,Lewenstein2007a,Ablowitz2009a,Lee2009a,Bermudez2010a,Chen2011a,Haddad2011a,Soltan-Panahi2011a,Tarruell2012a,Lim2012a,Jotzu2014a,Li2016a,Cooper2019a}.
A honeycomb supersolid is unique because its lattice structure emerges spontaneously and coexists with superfluidity. As a result of these multiple broken symmetries, a 2D supersolid is expected to exhibit three gapless excitation bands associated with Nambu-Goldstone modes \cite{Watanabe2012a}.

{In this Letter, we present the  results for the excitation spectrum of the honeycomb supersolid, building on prior work examining triangular 2D supersolid excitations \cite{Saccani2012a,Watanabe2012a,Kunimi2012a,Macri2013a,Poli2024b} and honeycomb ground state and dynamical properties \cite{Zhang2019a,Ripley2023a,Zhang2024a}}. The three lowest excitation bands can be characterized as longitudinal first and second sound, as well as a transverse sound. These excitations are dominated by the incompressible nature of the dipolar system, causing the speed of first sound (crystal modes) to be much higher than the other two sounds.  Our results show that Dirac points emerge in the lowest two sound branches.

\begin{figure}[htbp!]
	\centering
	\includegraphics[width=3.2in]{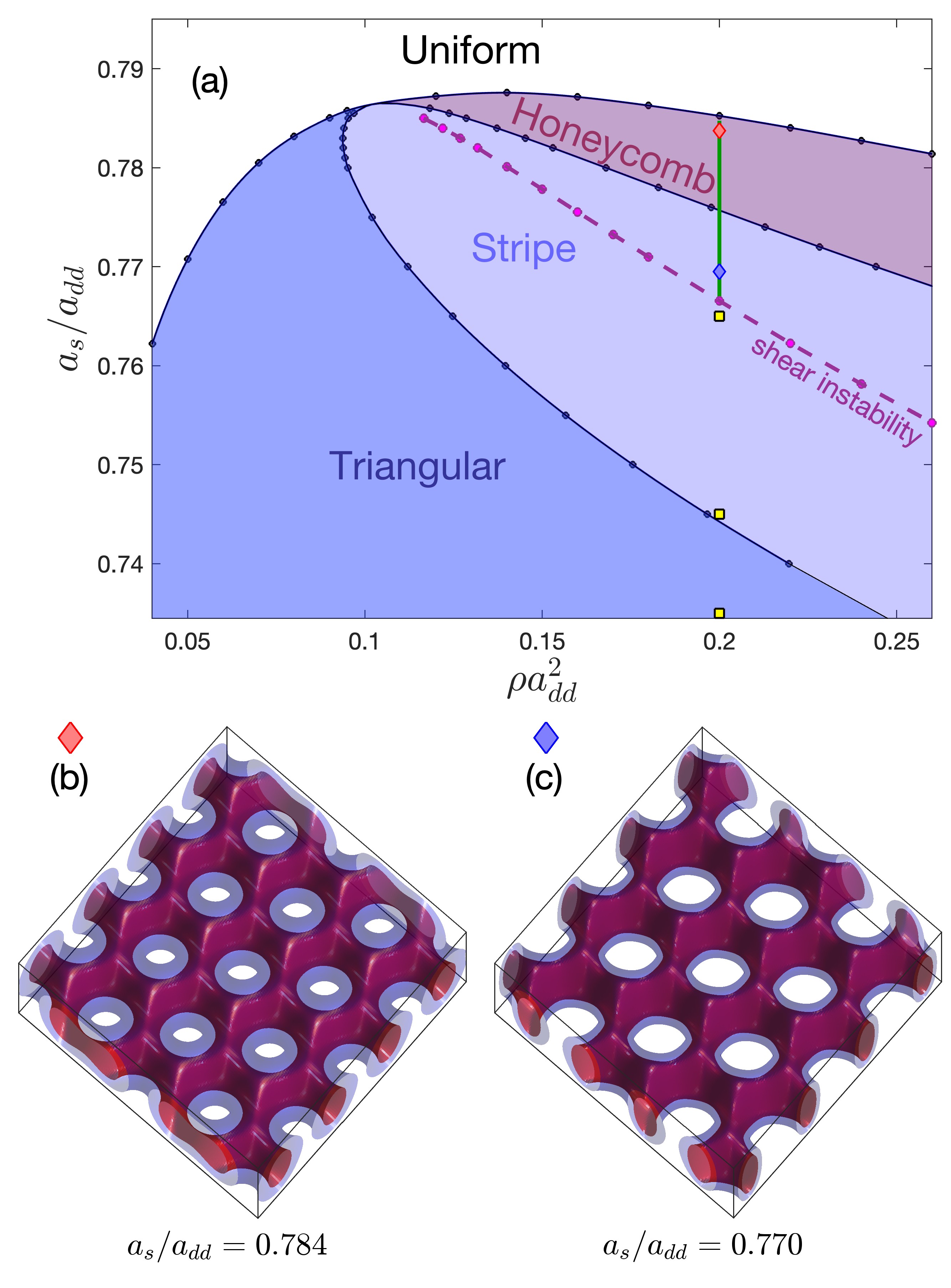} 
	\caption{(a) Ground state phase diagram for a planar dipolar BEC of $^{164}$Dy. Boundaries determined by finding where adjacent phases are degenerate (blue circles). Green line shows the parameter regime in Fig.~\ref{soundfig} and the dashed magenta line shows where the honeycomb state becomes unstable. Yellow squares indicate quench parameters [see Fig.~\ref{quenchdynamics}]. (b), (c) Example honeycomb ground states with parameters indicated by diamond markers in (a) and plotted in a cubic region of side length $20\,\mu$m. Red (blue) colored isosurfaces are at a density of $3\times10^{20}\,$m$^{-3}$ ($1.5\times10^{20}\,$m$^{-3}$).   
	\label{phasediag}}
\end{figure}

\begin{figure*}[htbp!]
	\centering 
	\includegraphics[width=7in]{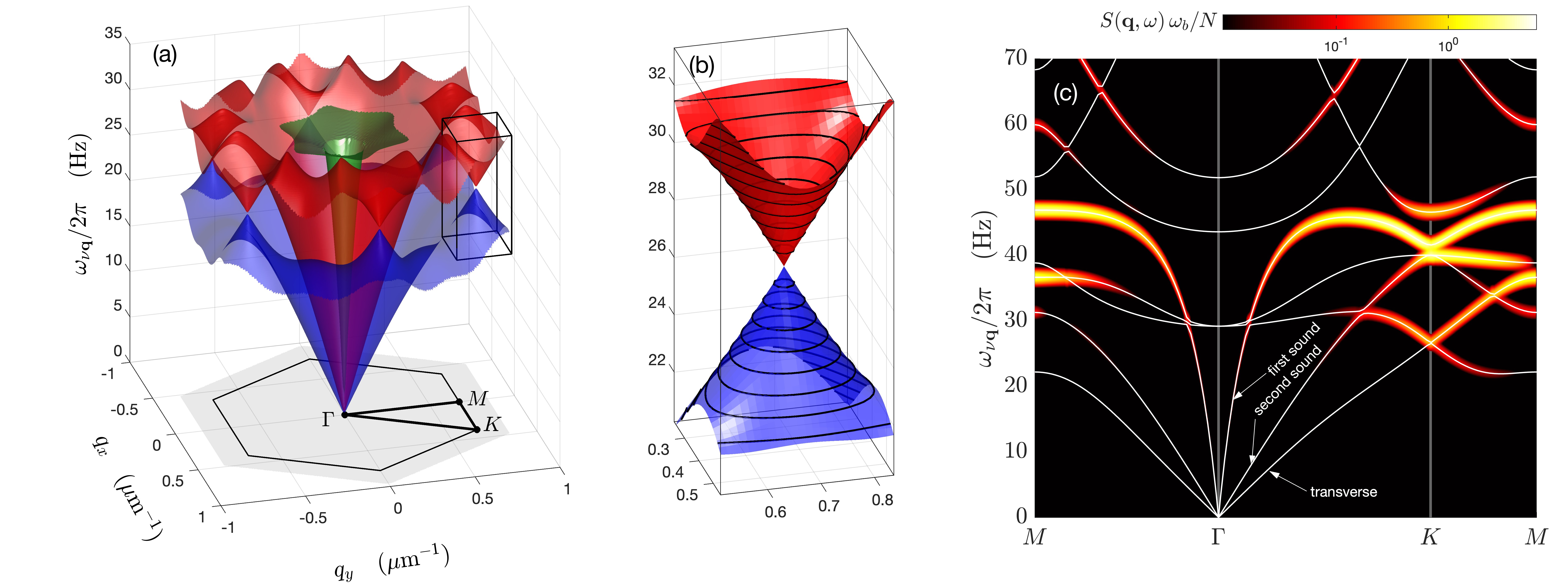}
	\caption{Band structure of a honeycomb supersolid. (a) Lowest three excitations bands.   The symmetry points $\{\Gamma,M,K\}$ and the first Brillouin zone are indicated for reference. (b) Close up of Dirac cones  near the $K$-point [range indicated by rectangular cuboid in (a)]. (c) Excitations along the three symmetry directions of the Brillouin zone: $\hbar\omega_{\nu\mathbf{q}}$  (white lines) on a heat map image of $S(\mathbf{q},\omega)$ [frequency broadened by $\omega_b=10^{-2}\omega_z$,  {with $N$ the atom number in the quantization volume (see Ref.~\cite{BECbook})]}.
	Results for   $\rho=0.2/a_{dd}^2$ and $a_s/a_{dd}=0.780$.	\label{bandstructure}}
\end{figure*}

We also explore the behavior of the sound speeds and elastic parameters as the contact interaction  ($a_s$)  is varied across the transition to the stripe-patterned supersolid phase. Here we find that the honeycomb state remains metastable until the shear modulus (and transverse speed of sound) eventually vanishes, signalling the loss of crystal rigidity. These conclusions are validated through non-equilibrium simulations, demonstrating this instability can cause a transition to states with stripe, triangular and mixed crystalline order. Intriguingly, phase coherence can persist during these crystal transitions.

\noindent{\bf Formalism} -- We present results for the case of a dipolar BEC of highly magnetic $^{164}$Dy atoms with magnetic dipole moments polarized along $z$ and in a planar trap with axial harmonic confinement of angular frequency $\omega_z/2\pi=72.4\,$Hz. These atoms have a short ranged interaction with $s$-wave scattering length $a_s$ that can be adjusted with Feshbach resonances \cite{Baumann2014a}, and a long-ranged dipole-dipole interaction with dipole length $a_{dd}  = m\mu_0\mu_m^2/12\pi\hbar^2=130.8\,a_0$,  where $\mu_m=10\mu_B$  is the  magnetic moment and $a_0$ is the Bohr radius. 
The extended meanfield theory energy functional  is  
\begin{align}
E &= \int d\bx\, \psi^*\left(h_{\mathrm{sp}}+ \tfrac12\Phi   +\tfrac25\gammaQF|\psi|^3\right)\psi,\label{Efunc}
\end{align}
where $h_{\mathrm{sp}}=-\frac{\hbar^2}{2m} \nabla^2  + \frac12 m\omega_z^2z^2$ is the single particle Hamiltonian.
 Here $\Phi(\mathbf{x})=\int d\mathbf{x}^\prime\,U(\mathbf{x}-\mathbf{x}^\prime)|\psi(\mathbf{x}^\prime)|^2$ describes the two-body interactions, with the interaction potential
\begin{equation}
	U(\br) = \frac{4\pi a_s\hbar^2}{m}\delta(\br) + \frac{3\add\hbar^2}{m r^3}\left(1-3\frac{z^2}{r^2}\right).
\end{equation} 
Quantum fluctuation effects are important  when the  dipolar interactions  dominate  ($a_{s}/a_{dd}<1$) and stabilise the condensate from mechanical collapse \cite{Petrov2015a,Ferrier-Barbut2016a,Wachtler2016a,Bisset2016a,Chomaz2022R}. The quantum fluctuation term has coefficient $\gammaQF =  {128\hbar^2\sqrt{\pi a_s^5}}  \mathcal{Q}_5(a_{dd}/a_s)/3m$, where $\mathcal{Q}_5(x)=\Re\{\int_0^1 du[1+x(3u^2 - 1)]^{5/2}\}$ \cite{LHY1957,Lima2011a}.

\begin{figure}[htbp!]
	\centering
	\includegraphics[width=3.0in]{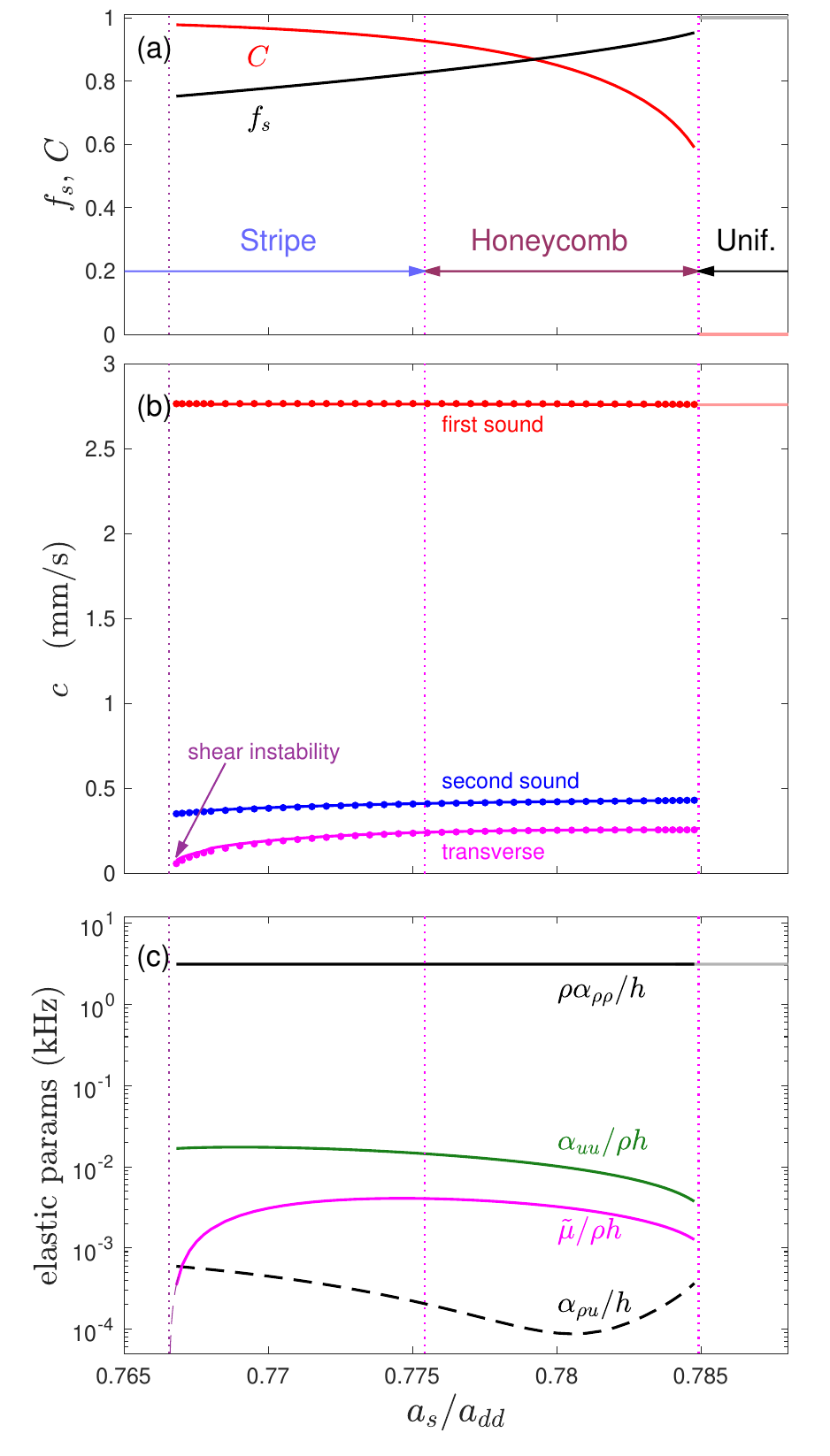} 
	\caption{Speeds of sounds and elastic parameters of honeycomb supersolid.  (a)  {Superfluid fraction $f_s$ and the planar density contrast $\mathcal{C}$  of the honeycomb state. Also shown for the uniform state where it is the ground state (lightly colored lines)}.  Labels indicate where the uniform, honeycomb and stripe are the ground states with vertical dotted lines indicating the transitions.
	(b) Speeds of sound from BdG calculations (lines) and hydrodynamic result (dots). Uniform superfluid result (light solid line). 
(c) Elastic parameters (solid and dashed lines), with $\alpha_{\rho\rho}$ for the uniform state (light solid line). Polynomial extrapolation determining instability point (dashed magenta line) where $c_t,\tilde{\mu}\to0$. 
The superfluid fraction is isotropic and calculated as in Ref.~\cite{Blakie2024a}, and   $\mathcal{C}=(\varrho_{\max}-\varrho_{\min})/(\varrho_{\max}+\varrho_{\min})$, where $\varrho_{\max}$ ($\varrho_{\min}$) is the maximum (minimum) of the areal density. 
The  parameters for these results are the green line in Fig.~\ref{phasediag}(a).   
	\label{soundfig}}
\end{figure}

\noindent{\bf Phase diagram} -- The phase diagram presented in Fig.~\ref{phasediag}(a) characterizes the ground state as a function of average areal density $\rho$ and interaction parameter ratio $a_s/a_{dd}$. This is determined by solving for the stationary states of the energy functional, which satisfy the extended Gross-Pitaevskii equation (eGPE) $\mu\psi=\mathcal{L}\psi$, where
$
\mathcal{L}\equiv h_{\mathrm{sp}} + \Phi   +\gammaQF|\psi|^3,
$
 and $\mu$ is the chemical potential. The average density constraint is enforced through the normalization condition $\int_{\mathrm{uc}} d\mathbf{x}\,|\psi(\mathbf{x})|^2=\rho A_{\mathrm{uc}}$, where $A_{\mathrm{uc}}$ is the unit cell area and the integration is taken over a unit cell of the periodic density pattern.
Depending on the parameters the condensate can prefer to be in a uniform superfluid state or a spatially modulated state of various crystal structures (see  Refs.~\cite{Zhang2019a,Lee2021a,Ripley2023a,Poli2024b,Smith2023a} for details of these calculations).  We show two examples of honeycomb ground states in Figs.~\ref{phasediag}(b) and (c). The connected ``rings" of the honeycomb state confer a high superfluid fraction \cite{Zhang2019a,Ripley2023a,Blakie2024a} relative to the triangular ground state (and related one-dimensional dipolar supersolids), where the unit cell is dominated by localized droplet-like states, and tunnelling to neighboring sites is highly suppressed.
 {The transitions between states are generally first-order, e.g.~see the  Fig.~\ref{soundfig}(a), where an abrupt change in the contrast (density modulation) and the superfluid fraction occurs at the uniform to honeycomb transition point.}

\noindent{\bf Excitation spectrum} -- The collective excitations of the supersolid are determined by solving the extended Bogoliubov-de Gennes (BdG) equations  \cite{Baillie2017a}: $H\mathbf{w}_{\nu\mathbf{q}}=\hbar\omega_{\nu\mathbf{q}}\mathbf{w}_{\nu\mathbf{q}}$, where
\begin{align}
H=   \left[\begin{matrix} 
      \mathcal{L}+X-\mu & -X \\
      X & -(\mathcal{L}+X-\mu) \\
   \end{matrix}\right],
\end{align}
with $\mathbf{w}_{\nu\mathbf{q}}=[u_{\nu\mathbf{q}},v_{\nu\mathbf{q}}]^T$ and $\{\hbar\omega_{\nu \mathbf{q}}\}$ being the excitation modes and energies, respectively. Here $\mathbf{q}=(q_x,q_y)$ is the quasimomentum, $\nu$ is the band index and 
\begin{align}
\!Xf\!=\!\psi(\mathbf{x})\!\int\!d\mathbf{x}^\prime U(\mathbf{x}\!-\!\mathbf{x}^\prime)f(\mathbf{x}^\prime)\psi^*(\mathbf{x}^\prime) +\frac32\gamma_{\mathrm{QF}}|\psi|^3f.
\end{align}

We give results for the honeycomb excitation spectrum in Fig.~\ref{bandstructure}.
As expected for a 2D supersolid, the lowest three-excitation bands are gapless \cite{Watanabe2012a}.  These arise from the Nambu-Goldstone modes associated with broken phase symmetry and the broken  2D translational symmetry. The lowest excitation band is a transverse excitation, and the higher two are longitudinal excitations referred to first and second sound [see Fig.~\ref{bandstructure}(c)]. Second sound manifests from the normal fraction arising from the broken translational invariance \cite{Leggett1998a}. The lowest three bands are shown over the first Brillouin zone in Fig.~\ref{bandstructure}(a). The band structure is observed to have 6 Dirac points at the edge of the Brillouin zone. At these points the lowest two bands are degenerate and close by the dispersion of these bands is linear (cone like) [see Fig.~\ref{bandstructure}(b)].  

The band structure along the symmetry lines is shown in Fig.~\ref{bandstructure}(c). Here we also display the dynamic structure factor
\begin{align}
S(\mathbf{q},\omega)=\sum_{\nu}\left|\int d\mathbf{x}(u_{\nu \mathbf{q}}^*-v_{\nu \mathbf{q}}^*)e^{i\mathbf{q}\cdot\mathbf{x}}\psi\right|^2\delta(\omega-\omega_{\nu\mathbf{q}}),\label{Sqw}
\end{align}
which characterizes the density fluctuations in the system and demonstrates how strongly the various excitation branches respond to a density coupled probe \cite{BECbook}. 
Except near the Brillouin zone edges, most of the weight of $S(\mathbf{q},\omega)$ is from the first sound mode (which crosses from being the 3rd band near $\Gamma$ to be the fifth band near $M$ and $K$).  The lowest two bands near the Dirac points develop significant weight indicating that this feature will be susceptible to probing.

\noindent{\bf Speeds of sound}  -- The speeds of sound can be determined from the slope of the lowest three bands near the $\Gamma$ point. These are shown as function of  $a_s$ in Fig.~\ref{soundfig}(b).  These results take a slice through the phase diagram crossing into  the stripe state phase (i.e.~honeycomb is metastable) [see Fig.~\ref{phasediag}(a)]. In this metastable regime the transverse speed of sound softens to zero marking a shear instability.
We also note that the  first sound speed is much higher than the other sounds. A relatively high first sound speed is also found for one-dimensional and the triangular 2D dipolar supersolids \cite{Roccuzzo2019a,Blakie2023a,Sindik2024a,Poli2024b}.

\noindent{\bf Elastic parameters}   --
We can define elastic parameters from understanding how the ground state energy density $\mathcal{E}$  (i.e.~energy per area) changes with the average density $\rho$  and the lattice constants describing the crystal periodicity. This introduces the parameters $\{\alpha_{\rho\rho},\alpha_{uu},\alpha_{\rho u},\tilde{\mu}\}$ that appear in the superfluid hydrodynamic theory. The first  parameter,
$\alpha_{\rho\rho} \equiv \frac{\partial ^2\mathcal{E}}{\partial \rho^2}$, 
relates to the isothermal compressibility at constant strain $\tilde\kappa=({\rho^2\alpha_{\rho\rho}})^{-1}$. 
The honeycomb lattice has an isotropic elastic tensor  \cite{LandauElasticity} characterized by  the Lam\'e parameters $\{\tilde\lambda,\tilde\mu\}$, with $\tilde\mu$ being the shear modulus. The longitudinal modulus, $\alpha_{uu}\equiv\tilde\lambda+2\tilde\mu$,   is the diagonal element of the elastic tensor. We also evaluate the density-strain coupling parameter $  \alpha_{\rho u}$, which quantifies the mixed derivative of  $\mathcal{E}$ with respect to changes in density and cell area, although this tends to be small. The elastic parameters are calculated from ground state solutions using finite differences in $\rho$ and unit cell distortion, and are shown in Fig.~\ref{soundfig}(c) (e.g.~see Refs.~\cite{Platt2024a,Poli2024b,Rakic2024a}).

\begin{figure*}[htbp!]
	\centering 
	\includegraphics[width=7in]{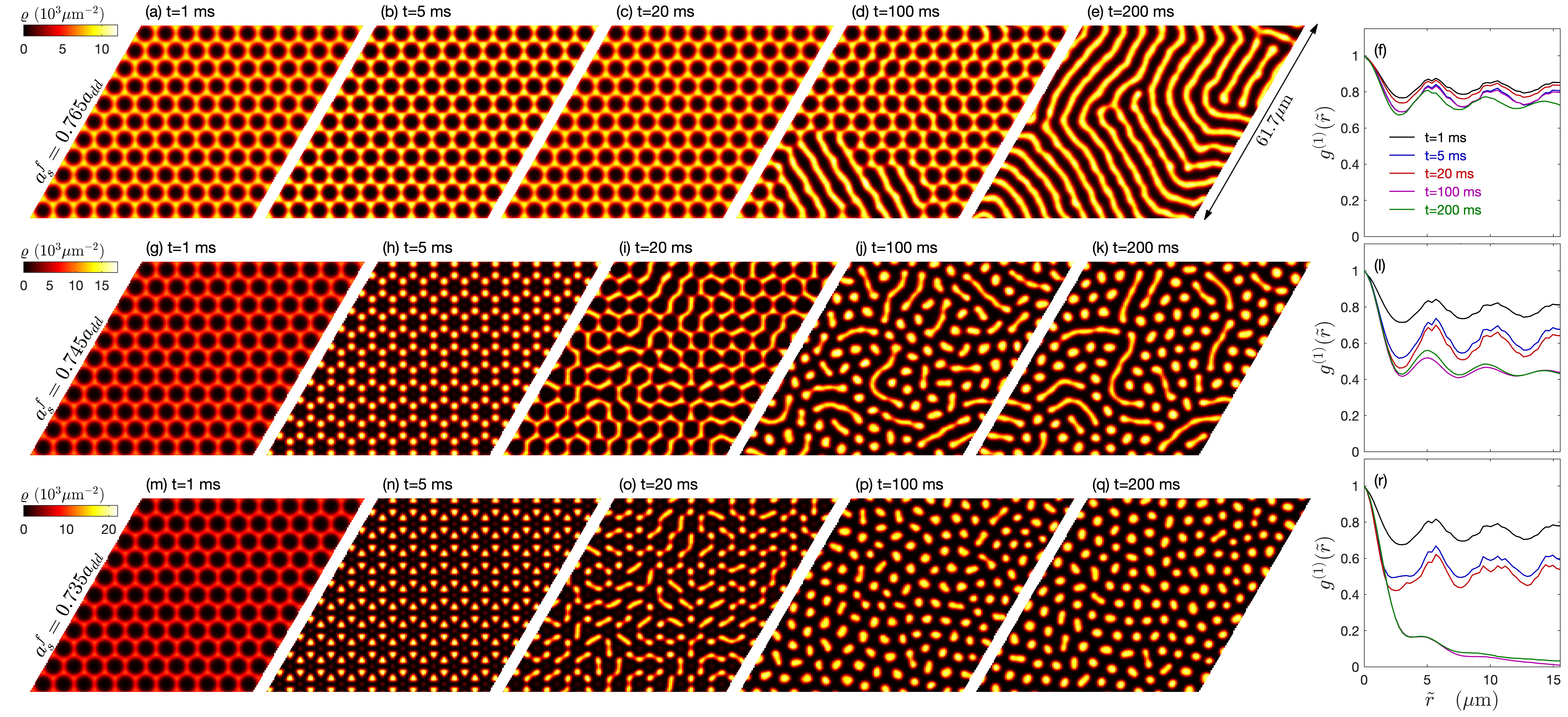}
	\caption{Dynamics of a honeycomb supersolid following a quench into the unstable regime.  
	Areal density $\varrho$  and coherence evolution for quenches to  (a)-(f) $a_s^f=0.765a_{dd}$, (g)-(l) $a_s^f=0.745a_{dd}$, and (m)-(r)  $a_s^f=0.735a_{dd}$ [final parameters shown on Fig.~\ref{phasediag}(a)]. Initial state for  $a_s^i=0.770\,a_{dd}$   [Fig.~\ref{phasediag}(c)] with noise added.  Coherence function  $g^{(1)}(\tilde{r})=G(\tilde{r})/G(0)$, where $G(\tilde{r})$ is the angular average of the  correlation function $G(\tilde{\mathbf{r}})=\frac{1}{A}\int d\tilde{\mathbf{r}}^\prime\,\psi^*(\tilde{\mathbf{r}}^\prime+\tilde{\mathbf{r}})\psi(\tilde{\mathbf{r}}^\prime)$, with $\tilde{\mathbf{r}}^\prime=(x,y,0)$ and $A$ being the system area. Other parameters as in Fig.~\ref{phasediag}.  
	\label{quenchdynamics}}
\end{figure*}

 The hydrodynamic theory \cite{Yoo2010a,Platt2024a,Poli2024b} relates the speeds of sound to the elastic parameters as
\begin{align}
    mc_\pm^2= & \varepsilon  \pm \sqrt{ {\varepsilon^2} \!-\!  \frac{\rho_{s}}{\rho_{n}}\left(\alpha_{\rho\rho}\alpha_{uu}-\alpha_{\rho u}^2\right)},\quad
  c_t=\sqrt{\frac{\tilde{\mu}}{\rho_{{n}}}},\label{speedsofsound}
    \end{align} 
  being first ($c_+$), second ($c_-$) and transverse ($c_t)$ sound, respectively,
where we have set $\varepsilon  =  (\rho\alpha_{\rho\rho}-2\alpha_{\rho u} +  {\alpha_{uu}}/{\rho_n})/2$. Here $\rho_n=(1-f_s)\rho$  and $\rho_s=f_s\rho$  are the normal and superfluid density, respectively, where $f_s$ is the superfluid fraction.
The speeds of sound determined by the BdG and hydrodynamic calculations are in good agreement [Fig.~\ref{soundfig}(b)].  As expected from Eq.~(\ref{speedsofsound}), $c_t\to0$ at the shear instability due to $\tilde{\mu}$  vanishing. This signals the honeycomb lattice losing rigidity.  Extending this analysis to other densities we determine the instability boundary on the phase diagram [Fig.~\ref{phasediag}(a)].   Above this line the honeycomb state is dynamically stable. It would be interesting to extend this line to the critical point, where all 4 phases meet, however the density contrast vanishes as we approach the critical point and a careful analysis of the shear instability in this region is difficult.

\noindent{\bf Quench dynamics}  -- We explore the metastability  and instabilities of the honeycomb supersolid with the dynamics following an $a_s$ quench. We perform simulations of finite sample of $11\times11$ cells (system in a right rhombic prism shaped spatial region) with periodic boundary conditions on the  $xy$-plane boundaries. The initial state is the metastable honeycomb state at $a_s^{i}/a_{dd}= 0.770$ [Fig.~\ref{phasediag}(c)] with white noise added (increasing the energy by $\sim5\%$) to simulate the effects of quantum and thermal fluctuations, which play in important role in seeding instabilities \cite{cfieldRev2008}.  
This initial state is evolved using the truncated Wigner method (see \cite{cfieldRev2008,Bisset2015a})  with $a_s$ quenched to the final value $a_s^{f}$ at  $t=0$. We present three example trajectories in Fig.~\ref{quenchdynamics} for different values of $a_s^{f}$. Generically the quench excites a breathing oscillation where the peak density and width of the honeycomb walls oscillate with a period of $\sim10\,$ms [e.g.~see Figs.~\ref{quenchdynamics}(a) and (b)]. If  $a_s^f$ is above the shear instability,  the oscillation continues for the duration of simulation without any degradation of the honeycomb structure. For values of $a_s^f$ below the instability, the honeycomb structure breaks down and the system reorganizes. For the quench in Figs.~\ref{quenchdynamics}(a)-(e) a disordered stripe-like pattern starts forming at $t\sim100\,$ms. This decay happens more rapidly for lower  $a_s^f$, but also the final pattern changes. For values of $a_s^f$ in the triangular region of the phase diagram [Figs.~\ref{quenchdynamics} (m)-(q)], a  triangular pattern emerges, and for a value between these two cases a mixed stripe-triangular arrangement develops [Figs.~\ref{quenchdynamics} (g)-(k)]  {(cf.~intriguing metastable states  found in the middle of the stripe phase in Ref.~\cite{Zhang2024a}).} In subplots (f), (l) and (r) we consider the evolution of the phase coherence. This reveals that final states with some stripe component  can maintain a high phase coherence  ($\gtrsim50\%$), and can be expected to exhibit a high superfluid rigidity.

\noindent{\bf Outlook and Conclusions} --
In this paper, we have analyzed the excitations of a honeycomb supersolid revealing  the general behavior of sound, and the occurrence of Dirac points. Schemes for performing band spectroscopy on dipolar supersolids have been proposed and implemented in experiments \cite{Petter2021a,Sindik2024a,Biagioni2024a}  and our results show that the Dirac cones will respond to such density-coupled probes.
Considering the behavior across the phase diagram, we observe that the honeycomb state is metastable over a significant portion of the stripe phase region  {\cite{Zhang2024a}} until a shear instability manifests. This can be compared to the vanishing of shear associated with the (finite-temperature) melting of a solid-state crystal \cite{Born1940a}.  
 {Also, it has been found that thermal effects can modify the dipolar supersolid phase diagram \cite{Sanchez-Baena2023a,Sanchez-Baena2024a}, suggesting  the extension of our study to finite temperature is of interest.}

We have presented non-equilibrium simulations exploiting this instability, showing how $a_s$ quenches lead to the honeycomb crystal melting into stripe, triangular and mixed phases on time-scales accessible to experiments. These types of quenches will be suitable for mapping out the phase diagram, verifying metastability, and open new directions for studying nonequilibrium dynamics of transitions between different types of supersolids.   The planar results we consider provide a general understanding of dipolar BECs prepared in pancake-shaped traps, but  will be more closely realized in experiments employing optical box traps  \cite{Juhasz2022a}.

\noindent{\bf Acknowledgments} --
The New Zealand eScience Infrastructure (NeSI)  and support from the Marsden Fund of the Royal Society of New Zealand. 

%

\end{document}